\documentstyle[aps,12pt,manuscript]{revtex}
\begin{document}
\preprint{INJE--TP--95--8  }
\def\overlay#1#2{\setbox0=\hbox{#1}\setbox1=\hbox to \wd0{\hss #2\hss}#1%
\hskip -2\wd0\copy1}

\title{ Two-dimensional extremal black holes}

\author{ H.W. Lee and Y. S. Myung }
\address{Department of Physics, Inje University, Kimhae 621-749, Korea}
\author{Jin Young Kim}
\address{Division of Basic Science, Dongseo University, Pusan 616-010, Korea}

\maketitle
\vskip 1.5in

\begin{abstract}
We discuss the  two-dimensional (2D), $\epsilon<2$ extremal ground states of
 charged black hole. Here $\epsilon$ is  the dilaton coupling parameter for the
Maxwell term.
The complete analysis of stability is carried out for all these extremal black
holes.
It is found that they  are all  unstable. To understand this instability, we
study
the non-extremal charged black hole with two (inner and outer) horizons. The
extremal black holes
appear when two horizons coalesce. It conjectures that the instability
originates from
the inner horizon.
\end{abstract}

\newpage
\section{Introduction}
Recently the extremal black holes have received much attention.  Extremal black
holes provide
a simple laboratory in which to investigate the quantum aspects of black hole
[1]. One of the crucial
features is that the Hawking temperature vanishes.  The black hole with $M>Q$
  will radiate down to its extremal $M=Q$ state. Thus the extremal black hole
may play  a role of
the stable endpoint for the Hawking evaporation. It has been also proposed that
although
the extremal black hole has nonzero area, it has zero entropy [2]. This is
because the extremal case is
distinct topologically  from the nonextremal one.

It is very important to inquire into the stability of the extremal black holes,
which  is essential
to establish their
physical existence. We note that  the 4D extremal  charged black holes with the
coupling parameter
($a$) are shown to be classically  stable [3]. The $a=0$ case corresponds to
the extremal
Reissner-Nordstr\"om black hole.  Since all potentials are positive definite
outside the
 horizon, one can easily infer  the stability of 4D extremal  charged black
holes  using the same
argument as employed by Chandrasekhar [4]. To the contrary,
it appears that the 2D  extremal black holes are unstable [5].

In this paper, we  consider the  prototype model for  extremal black holes.
This is the 2D, $\epsilon<2$ ($\epsilon$ is the 2D coupling parameter
corresponding to $a$)
 extremal black holes [6]. It is very important to find  whether  these black
holes are  stable or not.
For this end, one reminds that an extremal black hole is regarded as the limit
of a
non-extremal charged black hole. The 2D charged black hole has the event
(outer)
as well as the Cauchy (inner) horizons.
The stability of the outer horizon
is followed  by the conventional argument of the stability.
 One easy way of understanding  a black hole is to find
out how it reacts to external perturbations.
We always  visualize the
black hole as presenting an effective potential barrier (or well) to the
on-coming waves [4].
 As a  compact criterion for a single horizon (for an extremal black hole) or
outer horizon, the horizon is unstable if there
exists the potential well to the on-coming waves [7]. This is so because  the
Schr\"odinger-type equation
with the potential well always allows the bound state as well as scattering
states.
The former shows up as an imaginary frequency mode, leading to an exponentially
growing mode.
If one finds any exponentially growing mode, the extremal
black hole (or the outer horizon of non-extremal black hole) is unstable.
In Ref.[8] a conformally coupled scalar  ($f_i$) is used as a test field, while
in our case
a tachyon is used as a test field. If one takes a conformally coupled scalar to
study the black
hole, one finds the free field equation for the linearization : $\bar \nabla^2
f_i=0 \to
 (d^2/dr^{*2} +\omega^2)f_i=0$.
 This implies that one cannot find
the potentials, which are crucial for obtaining information about the black
holes.
 Although $f_i$ is useful for a semiclassical study of
the black hole, it is not a good  field for our study.

On the other hand, the stability analysis of the inner horizon is related to
how waves from the external
world (entering the inner region between two horizons via the outer one)
develop as they approach the
inner one and affect a free-falling observer (FFO). If a FFO crossing the inner
horizon
meets an infintely infalling radiation, one concludes that the inner horizon is
unstable.
This is a direct consequence of the non-trivial causal structure of the
space-time:
An observer crossing the inner horizon sees the entire history of the universe
in a tremendous flash.
We will show that the inner horizons of  the 2D, $\epsilon<2$ charged black
holes are unstable, whereas the
outer horizons are stable.
One finds  the barrier-well type potentials ($V^{IN}$) between
the Cauchy horizon and event horizon, while a potential barrier ($V^{OUT}$) is
induced
outside the event horizon. When these  coalesce (extremal case),
a barrier-well type potential persists  outside the event horizon.
This induces the instability of the extremal black hole.

The organization of this paper is as follows. We discuss the 2D, $\epsilon<2$
black hole solutions
in Sec.II.  Analyzing the metric function $f(r,\epsilon,Q)$,  one finds the
extremal, and non-extremal black holes.
For the stability analysis, in Sec.III we linearize the equation of motion
around the
 background solution. Sec. IV is devoted to investigate the stability for
extremal black holes.
It turns out that they are all unstable. In order to understand this
instability, we study
the non-extremal black holes in Sec. V. The inner structure of charged black
hole is explored
to understand the nature of the Cauchy horizon. The Hawking temperature is
calculated in Sec. VI.
 Finally we discuss our results in Sec. VII.

\section{ The $\epsilon<2$ black hole solution}
We  start with  two-dimensional dilaton gravity ($\Phi,G_{\mu\nu}$) conformally
coupled to Maxwell
($F_{\mu\nu}$) and tachyon ($T$)[10]
\begin{equation}
S_{l-e} = \int d^2 x \sqrt{-G} e^{-2\Phi}
   \big \{ R + 4 (\nabla \Phi)^2 +\alpha^2 - {1 \over 2}e^{\epsilon \Phi} F^2 -
{1 \over 2} (\nabla T)^2  +T^2\big \}.
\end{equation}
The above action with $\epsilon=0$ can be realized from the
heterotic string. We  introduce the tachyon to study the properties of black
holes in a simple way.
Setting $\alpha^2 = 8$ and  after deriving equations of motion, we take the
transformation
\begin{equation}
-2\Phi \rightarrow \Phi,~~~ T \rightarrow \sqrt 2 T, ~~~-R  \rightarrow R.
\end{equation}
Then the equations of motion become
\begin{eqnarray}
&&R_{\mu\nu} + \nabla_\mu \nabla_\nu \Phi  + \nabla_\mu T \nabla_\nu T
+{ 4 - \epsilon \over 4}e^{-\epsilon \Phi/ 2}  F_{\mu\rho}F_{\nu}^{~\rho} = 0,
\\
&& \nabla^2 \Phi + (\nabla \Phi)^2  - {1 \over 2}e^{-\epsilon \Phi/ 2} F^2  - 2
T^2 - 8 = 0,  \\
&&\nabla_\mu F^{\mu \nu} + {2- \epsilon \over 2} (\nabla_\mu \Phi) F^{\mu \nu}
= 0,   \\
&&\nabla^2 T + \nabla_\mu \Phi \nabla^\mu T + 2 T = 0.
\end{eqnarray}

An electrically  charged black hole solution to the above equations is obtained
as

\begin{equation}
\bar \Phi = 2 \sqrt 2 r,~~~ \bar F_{tr} = Q e^{-(2-\epsilon) \sqrt 2 r},~~~
\bar T = 0,
{}~~~ \bar G_{\mu\nu} =
 \left(  \begin{array}{cc} - f & 0  \\
                            0 & f^{-1}   \end{array}   \right),
\end{equation}
with the metric function
\begin{equation}
f = 1 -  {M \over \sqrt 2}e^{- 2 \sqrt 2 r} + {Q^2 \over 4(2- \epsilon)}e^{-
(4-\epsilon) \sqrt 2 r}.
\end{equation}
Here $M$ and $Q$ are the mass and electric charge of the black hole,
respectively.
We hereafter take $ M=\sqrt2$ for convenience. Note that from the requirement
of the finiteness
of electric energy ($\bar F(r\to \infty) \to 0$) and the asymptotically flat
spacetime
($f(r\to \infty) \to 1$), we have
the important constraint : $\epsilon <2$.

We analyze the metric function $f(r,\epsilon,Q)$ in (8) explicitly.
In general, from $f=0$ we can obtain
two roots ($r_{\pm}$) where $r_{+}(r_{-})$ correspond to the event (Cauchy)
horizons of
the charged black hole.   One is  interested  in the extremal limit
(multiple root: $r_+=r_- \equiv r_o$)
of the charged holes. This may provide a toy model to investigate the late
stages
of Hawking evaporation.
First of all, we wish to obtain the condition for $f(r,\epsilon,Q)=0$ to have
the multiple root.
For $\epsilon<2$, the shape of $f(r,\epsilon,Q)$ is always concave ($\smile$).
Thus the multiple root is always obtained when $f=0$ and $f^\prime  = 0$,
which impliy that $Q^2(\epsilon)=Q_e^2(\epsilon)
=8({2-\epsilon \over 4-\epsilon})^{(4-\epsilon)/2}$.
Here the prime $(\prime)$ denotes the derivative with respect to $r$.
 The extremal horizon is located at
\begin{equation}
 r_o(\epsilon)= - {1 \over 2 \sqrt 2} \log ({ 4-\epsilon \over 2-\epsilon}).
\end{equation}
The explicit form of the extremal $f_e$ is
\begin{equation}
f_e(r,\epsilon) = 1 - e^{- 2 \sqrt 2 r} + {2 \over (2- \epsilon)} ({2-\epsilon
\over 4-\epsilon})
^{(4-\epsilon)/2}e^{- (4-\epsilon) \sqrt 2 r}.
\end{equation}
Fig. 1 shows  the shapes of $f(r,\epsilon,Q)$ for three cases : $Q^2<Q_e^2,
Q^2=Q_e^2$ and $Q^2> Q_e^2$
for $\epsilon =0.5$.
For $Q^2<Q_e^2$, we find the non-extremal black hole. One always obtains
two roots($r_{\pm}$) :$f(r_+,\epsilon,Q)= f(r_-,\epsilon,Q)=0$. The
analytic forms of $r_{\pm}$ are not known because of the complex form of
the metric function.  However the numerical values of $r_{\pm}$ can be obtained
for given $\epsilon$
and $Q$. Hence our information for the metric function $f(r,\epsilon,Q)$
comes mainly from
the graphical analysis.
In the case of $Q^2>Q_e^2$, there does not exist any solution to
$f(r,\epsilon,Q)=0$,
which means that there is no horizon.
Hereafter we no longer consider this case.

\section{Linear perturbation}

We introduce small perturbation fields  around
the background solution as [5]
\begin{eqnarray}
&&F_{tr} = \bar F_{tr} + {\cal F}_{tr} = \bar F_{tr} [1 - {{\cal F}(r,t) \over
Q}],        \\
&&\Phi = \bar \Phi + \phi(r,t),                       \\
&&G_{\mu\nu} = \bar G_{\mu\nu} + h_{\mu\nu}  = \bar G_{\mu\nu} [1 - h (r,t)],
  \\
&&T = \bar T + \widehat t \equiv \exp (-{\bar \Phi \over 2}) [ 0 + t (r,t) ].
\end{eqnarray}
 One has to linearize (3)-(6) in order to obtain the equations governing the
perturbations as
\begin{eqnarray}
&&  \delta R_{\mu\nu} (h)
+ \bar \nabla_\mu \bar \nabla_\nu \phi
- \delta \Gamma^\rho_{\mu\nu} (h) \bar \nabla_\rho \bar \Phi
+ {4- \epsilon \over 2} e^{-\epsilon \sqrt 2 r} \bar F_{\mu \rho} {\cal
F}_\nu^{~ \rho}
- {4- \epsilon \over 4} e^{-\epsilon \sqrt 2 r}\bar F_{\mu \rho} \bar
F_{\nu\alpha} h^{\rho \alpha}
\nonumber   \\
&&~~~~~~~~~~~~~~~~~~
- { \epsilon(4- \epsilon) \over 8} e^{-\epsilon \sqrt 2 r}\bar F_{\mu \rho}
\bar F_{\nu}~^{\rho} \phi
 = 0,
\end{eqnarray}

\begin{eqnarray}
&&  \bar \nabla^2 \phi
- h^{\mu\nu} \bar \nabla_\mu \bar \nabla_\nu \bar \Phi
- \bar G^{\mu\nu} \delta \Gamma^\rho_{\mu\nu} (h) \partial_\rho \bar \Phi
- h^{\mu\nu} \partial_\mu \bar \Phi \partial_\nu \bar \Phi
+ 2 \bar G^{\mu\nu} \partial_\mu \bar \Phi \partial_\nu \phi
- e^{-\epsilon \sqrt 2 r} \bar F_{\mu \nu} {\cal F}^{\mu \nu}
\nonumber   \\
&&~~~~~~~~~~~~~~~ + e^{-\epsilon \sqrt 2 r} \bar F_{\mu \nu}   \bar
F_{\rho}^{~\nu} h^{\mu\rho}
+ { \epsilon \over 4} e^{-\epsilon \sqrt 2 r} \bar F^2  \phi = 0,
\end{eqnarray}

\begin{equation}
  ( \bar \nabla_\mu + {2- \epsilon \over 2} \partial_\mu \bar \Phi )
    ( {\cal F}^{\mu \nu} - \bar F_\alpha^{~~\nu} h^{\alpha \mu}
                             - \bar F^{\mu}_{~~\beta} h^{\beta \nu} )
      + \bar F^{\mu \nu} (\delta \Gamma^\sigma_{\sigma \mu} (h)
      + {2- \epsilon \over 2} (\partial_\mu \phi))
 = 0,
\end{equation}

\begin{equation}
\bar \nabla^2 \widehat t + \bar \nabla_\mu \bar \Phi \bar \nabla^\mu \widehat t
+ 2 \widehat t = 0,
\end{equation}
where

\begin{eqnarray}
&&\delta R_{\mu\nu} (h) = {1 \over 2} \bar \nabla_\mu \bar \nabla_\nu
h^\rho_{~\rho}
 + {1 \over 2} \bar \nabla^\rho \bar \nabla_\rho h_{\mu\nu}
 - {1 \over 2} \bar \nabla^\rho \bar \nabla_\nu h_{\rho\mu}
 - {1 \over 2} \bar \nabla^\rho \bar \nabla_\mu h_{\nu\rho},   \\
&&\delta \Gamma^\rho_{\mu\nu} (h) = {1 \over 2} \bar G^{\rho\sigma}
( \bar \nabla_\nu h_{\mu\sigma} + \bar \nabla_\mu h_{\nu\sigma} - \bar
\nabla_\sigma h_{\mu\nu} ).
\end{eqnarray}

 From (17) one can express ${\cal F}$ in terms of $\phi$ and $h$ as

\begin{equation}
{\cal F} = - Q ( h + {2- \epsilon \over 2}\phi).
\end{equation}
This means that ${\cal F}$ is no longer an independent mode.
Also from the diagonal element of (15), we have

\begin{eqnarray}
&&  \bar \nabla^2 h - 2 \bar \nabla^2_t \phi - 2 \sqrt 2 \partial^r h
-(4-\epsilon) e^{-\epsilon
\sqrt 2 r} \bar F_{tr}{\cal F}^{tr} -{(4-\epsilon) \over 2} e^{-\epsilon
\sqrt 2 r} \bar F_{tr} \bar F^{tr} (h - {\epsilon \over 2}\phi)=0 ,    \\
&&  \bar \nabla^2 h - 2 \bar \nabla^2_r \phi + 2 \sqrt 2 \partial^r h
   -(4-\epsilon) e^{-\epsilon \sqrt 2 r} \bar F_{tr}{\cal F}^{tr}
-{(4-\epsilon) \over 2}
 e^{-\epsilon \sqrt 2 r} \bar F_{tr} \bar F^{tr} (h - {\epsilon \over
2}\phi)=0.
\end{eqnarray}
Adding the above two equations leads to

\begin{equation}
\bar \nabla^2 ( h - \phi) - { Q^2(4- \epsilon) \over 2}
     e^{ -(4-\epsilon) \sqrt 2 r} (h + {4- \epsilon \over 2} \phi) = 0.
\end{equation}
Also  the dilaton equation (16)  leads to
\begin{equation}
\bar \nabla^2 \phi + 4 \sqrt 2 f \phi' + 2 \sqrt 2 (f' + 2 \sqrt2 f) h
- {Q^2(4- \epsilon )\over 2} e^{ -(4-\epsilon) \sqrt 2 r}\phi = 0.
\end{equation}
And the off-diagonal element of (15) takes the form
\begin{equation}
 \partial_t \big\{ ( \partial_r  - \Gamma^t_{tr}) \phi + \sqrt 2  h  \big\}
   = 0,
\end{equation}
which provides us  the relation between $\phi$ and $h$ as
\begin{equation}
\phi' = -\sqrt2 h + {1 \over 2}{f' \over f} \phi + U(r).
\end{equation}
Here $U(r)$ is the residual gauge degrees of freedom and thus we set $U(r) = 0$
for simplicity.
Substituting (27) into (25), we have
\begin{equation}
\bar \nabla^2 \phi + 2 \sqrt 2 f' ( h + \phi)
- {Q^2(4- \epsilon) \over 2} e^{ -(4-\epsilon) \sqrt 2 r}\phi = 0.
\end{equation}
Calculating (24) + 2 $\times$ (28), one finds the other equation
\begin{equation}
\bar \nabla^2 (h + \phi) + 4 \sqrt 2  f' ( h + \phi)
- 2 Q^2 e^{ -(4-\epsilon) \sqrt 2 r} \{( h + 4 \phi) - { \epsilon \over 4} ( h
+ { 12- \epsilon
\over 2}\phi ) \} = 0.
\end{equation}
Although (24) and (29) look like the very complicated forms,
these reduce to
\begin{eqnarray}
&&\bar \nabla^2 ( h - \phi) = 0, \\
&&\bar \nabla^2 (h + \phi) + 4 \sqrt 2 f' ( h + \phi) =0
\end{eqnarray}
in the asymptotically flat region ($r \to \infty$).
This suggests that one  obtains  two graviton-dilaton modes.
However, it is important to check whether the graviton ($h$),  dilaton
($\phi$), Maxwell
mode (${\cal F}$)  and tachyon ($t$) are  physically propagating modes
in the 2D charged black hole background.
We review the conventional counting of degrees of freedom.
The number of degrees of freedom for the gravitational field ($h_{\mu\nu}$) in
$D$-dimensions is $(1/2) D (D -3)$.  For a Schwarzschild black hole,
we obtain two degrees of freedom. These correspond to the Regge-Wheeler mode
for odd-parity perturbation
and Zerilli mode for even-parity perturbation [4].  We have $-1$ for $D=2$.
This means that in
two dimensions
the contribution of the graviton is equal and opposite to that of a spinless
particle (dilaton).
The graviton-dilaton modes ($h+\phi, h-\phi$) are gauge degrees of freedom and
thus turn out to be
nonpropagating modes [6].
In addition, the Maxwell field has $D-2$ physical degrees of freedom.
The Maxwell field has no physical degrees of freedom for $D=2$. Actually from
(21) it turns out to be
a redundant one.
Since all these are  nonpropagating modes, it is  necessary to consider the
remaining one (18).
The tachyon as a spectator is a physically propagating mode.
This is used to illustrate many of the qualitative results about the 2D charged
black hole
in a simpler context.
Its linearized equation is
\begin{equation}
f^2 \partial^2_r t  + f\partial_r f \partial_r t - \{ \sqrt 2 f \partial_rf -2
f (1 - f) \}t  - \partial_t^2 t = 0.
\end{equation}
To study the physical implications,  the above equation
should be transformed into  one-dimensional Schr\"odinger-type equation.
Introducing a tortoise coordinate
$$r\to r^* \equiv g(r),$$
(32) can be rewritten as
\begin{equation}
f^2 g'^2 {\partial^2 \over \partial r^{*2}} t  + f ( f g'' +  f' g')
{\partial \over \partial r^* }t - \{ \sqrt 2 ff' - 2 f (1 - f) \} t
 - {\partial^2 \over \partial t^2} t = 0.
\end{equation}
Requiring that the coefficient of the linear derivative vanish, one finds the
relation
\begin{equation}
g' =  {1 \over f}.
\end{equation}
Assuming $t_\omega( r^*,t ) \sim \tilde t_\omega ( r^* ) e^{i\omega t}$,
one can cast (33) into the Schr\"odinger-type equation

\begin{equation}
\{ {d^2 \over dr^{*2}} + \omega^2 - V(r,\epsilon,Q)\} \tilde t_\omega = 0,
\end{equation}
where the effective potential $V(r,\epsilon,Q)$  is given by
\begin{equation}
V(r,\epsilon,Q) = f(\sqrt 2 f' - 2  (1 - f)).
\end{equation}

\section{ The extremal case}

First we consider the extremal black holes.
The potentials surrounding the extremal black holes are given by
\begin{equation}
V^e(r,\epsilon)  = 2 e^{- 2 \sqrt 2 r} f_e (r,\epsilon) \{ 1- 2({3-\epsilon
\over 2-\epsilon})
({2-\epsilon \over 4-\epsilon})^{(4-\epsilon)/2}e^{- (2-\epsilon) \sqrt 2 r}\}.
\end{equation}
After a concrete analysis, one  finds the barrier-well type potentials for
$\epsilon <2$.
For examples, Fig. 2 shows the shapes of potentials for $\epsilon $= 1.9, 0.5,
and $-3$.
In this case the roots of $V^e=0$ are $r=r_o, r_b,$ and $\infty$ in sequence.
Here the extremal horizon ($r=r_o$) comes from $f_e=0$ and $r= r_b,\infty$
from $ \sqrt 2 f_e' - 2 (1 - f_e)=0$.
Now let us translate the potential $V^e(r,\epsilon)$ into $V^e(r^*, \epsilon)$.
{}From (10) and (34), one can find the form of $r^*=g= \int^r dr/ f_e$.
Setting  $y= e^{-2 \sqrt 2 r}$, we integrate this as
\begin{equation}
r^*= r + {1 \over 2 \sqrt 2(4-\epsilon)} \log |f_e|
- {2-\epsilon \over 2 \sqrt 2 (4- \epsilon)} \int ^y {dy \over { 1- y +A y^{1+
B}}}
\end{equation}
with $A= {2 \over (2- \epsilon)} ({2-\epsilon \over 4-\epsilon})
^{(4-\epsilon)/2}$ and $B = 1- {\epsilon \over 2}$.
Since both the forms of $V^e(r,\epsilon)$ and $r^*$ are very complicated, we
are far from obtaining
the exact form of $V^e(r^*,\epsilon)$. But one can obtain  the approximate
forms
to $V^e(r^*,\epsilon)$  near the both ends.
Since the second and last terms in (38) approach zero as $r\to \infty$,
one finds  that
\begin{equation}
r^* \simeq r.
\end{equation}
Then (37) takes
the asymptotic form
\begin{equation}
V^e_{r*\to \infty}(r^*) \simeq 2 \exp( -2 \sqrt 2 r^*),
\end{equation}
which is independent of $\epsilon$.
On the other hand, near the horizon ($r=r_o$)  the last term in (38) completely
dominates over
the first two terms. Expanding $1- y +A y^{1+ B}$ in a Taylor series  about
$y_0=\exp(-2 \sqrt 2 r_o)$ leads to  ${ AB(1-B) \over 2}y_o^{B-1}(y-y_o)^2$.
Plugging this  into (38),  one has
\begin{equation}
r^* \simeq - { 2 \over \sqrt 2 (2-\epsilon)} { 1 \over ({4-\epsilon \over
2-\epsilon}-
  e^{ - 2 \sqrt 2 r})}.
\end{equation}
{}From (41), one finds the asymptotic relation
\begin{equation}
e^{- 2\sqrt 2 r} \simeq {4-\epsilon \over 2-\epsilon} + {2 \over \sqrt 2
(2-\epsilon)}{1 \over r^*}.
\end{equation}
Substituting this into (37) leads to the potential near the horizon $(r\to r_o,
r^* \to -\infty)$
\begin{equation}
V^e_{r*\to -\infty}(r^*,\epsilon) \simeq - {1 \over (4-\epsilon)}{ 1 \over
r^{*2}}.
\end{equation}
Using (40) and (43) one can construct the approximate form $V^e_{app}(r^*,
\epsilon)$ (Fig. 3).
This is also a  barrier-well which is  localized at the origin of $r^*$.
Our  analysis is  based on the approximate equation,
\begin{equation}
\{ {d^2 \over dr^{*2}} + \omega^2 - V^e_{app}(r^*,\epsilon)\} \tilde t_\omega^e
= 0.
\end{equation}
As is well known, two kinds of solutions to Schr\"odinger-type equation with
potential well
correspond to the bound and scattering states. In our case
$V^e_{app}(r^*,\epsilon)$ admits  two solutions  depending on
the sign of $\omega^2$ :
(i) For $\omega^2>0(\omega=$ real), the asymptotic solution for $\tilde
t_\omega^e$ is given by
$\tilde t^e_{\omega,\infty}  =  \exp(i \omega r^*)  + R \exp(- i \omega r^*),
r^*  \to \infty $ and
$\tilde t^e_{\omega,EH} = T\exp( i\omega r^*), r^*  \to - \infty $.
Here $R$ and $T$ are the scattering amplitudes of two waves which are
reflected and transmitted by the potential $V^e_{app}(r^*,\epsilon)$, when a
tachyonic wave of unit
amplitude with the frequency $\omega$ is incident on the black hole from
infinity.
\noindent $(ii)$ For $\omega^2<0$ ($\omega =-i \alpha$, $\alpha$ is positive
and real),
 we have the bound state.
Eq. (44) is given by

\begin{equation}
{d^2 \over d r^{*2}}\tilde t^e  =  (\alpha^2 + V_{app}(r^*)) \tilde t^e.
\end{equation}
The asymptotic solution is $\tilde t^e_\infty  \sim  \exp(\pm \alpha r^*), r^*
\to \infty $
and $\tilde t^e_{EH}    \sim  \exp(\pm \alpha r^*), r^*  \to - \infty $.
To ensure that the perturbation falls off to zero for large $r^*$, we choose
$\tilde t^e_\infty \sim \exp (-\alpha r^*)$.  In the case of $\tilde t^e_{EH}$,
the solution
$\exp (\alpha r^*)$ goes to zero as $r^* \to - \infty$.
Now let us observe whether or not $\tilde t^e_{EH} \sim \exp (\alpha r^*)$ can
be matched
to $\tilde t^e_\infty \sim \exp (-\alpha r^*)$.
Assuming $\tilde t^e$ to be positive,
the sign of $d^2 \tilde t^e / dr^{*2}$
can be changed from $+$ to $-$ as  $r^*$
goes from $\infty$ to $-\infty$.
If we are to connect $\tilde t^e_{EH}$ at one end to a decreasing solution
$\tilde t^e_\infty$
at the other, there must be a point ($d^2\tilde t^e/ dr^{*2}<0$,
$d \tilde t^e/dr^*=0$) at which the signs of $\tilde t^e$ and
$d^2\tilde t^e/dr^{*2}$ are opposite : this is  compatible with  the shape of
$V^e_{app}(r^*,\epsilon)$ in Fig. 3. Thus it is possible for
$\tilde t^e_{EH}$ to be connected to $\tilde t^e_\infty$ smoothly.  Therefore a
bound state solution
is given by

\begin{eqnarray}
\tilde t^e_\infty & \sim & \exp(- \alpha r^*),~~~~~~~~ ( r^*  \to \infty )
\\
\tilde t^e_{EH}   & \sim & \exp( \alpha r^*), ~~~~~~~~ ( r^*  \to - \infty ).
\end{eqnarray}
This is a regular solution everywhere in space at the initial time $t=0$.
It is well-known that in quantum mechanics, the bound state solution is always
allowed if
there exists a potential well.
The time evolution of the solution with $\omega=-i\alpha$ implies
$t^e_\infty (r^*, t)=\tilde t^e_\infty(r^*)\exp(-i\omega t) \sim \exp (-\alpha
r^*) \exp (\alpha t)$ and
$t^e_{EH} (r^*,t)=\tilde t^e_{EH}(r^*) \exp(-i\omega t) \sim \exp (\alpha r^*)
\exp (\alpha t)$.
This means that there exists an exponentially growing mode with time.
Therefore, the $\epsilon<2$ extremal ground states  are
classically unstable.

\section{The non-extremal case}

The origin of this instability comes from  the  barrier-well potentials.
These potentials  appear  in all $\epsilon<2$ extremal black holes.
This potential persisits when the non-extremal black hole approaches the
extremal limit (see Fig. 4).
An extremal black hole is considered as the limit of a
non-extremal one. A non-extremal black hole has an outer (event) and an inner
(Cauchy) horizon,
and these come together in the extremal limit.  In this sense, it is  necessary
to investigate
the non-extremal black holes.
The potential of the non-extremal black hole takes a barrier-well type between
the inner and
 outer horizons,  while it takes a simple barrier outside the outer horizon.

In this case the roots of $V=0$ are $r=r_-, r_b, r_+,$ in sequence.
Here the inner and outer horizons($r=r_-,r_+$) come from $f=0,$ and while $r=
r_b$
from $ \sqrt 2 f' - 2 (1 - f)=0$.
As is shown in Fig. 5, $V(r,\epsilon,Q)$ for $Q^2= Q^2_e/2$ increases without
bound in height
as $\epsilon$ approaches to 2 (the upper limit). The graphical analysis in Fig.
6 and 7 shows
that $V(r,\epsilon,Q)$ for $Q^2= Q^2_e/2$ are shifted to $V$-axis as $\epsilon$
decreases.
However, the shape of a barrier-well is not significantly changed under the
variation of $\epsilon$.

In this section we study mainly the inner structure of the $\epsilon<2$ charged
black holes.
It was shown that for $\epsilon=0$ the inner horizon is unstable, whereas the
outer one is stable [9].
First we consider the region inside the black hole ($r_-<r<r_+$).  Fig. 8 shows
that the inner structure
of the non-extremal black hole has the nontrivial causal stucture of the
spacetime.
 It is very important to note that inside the black
 hole the radial coordinate ($r$ or $r^*$) is timelike, whereas the time ($t$)
is spacelike.
 Hence to quest the internal structure of black hole is an evolutionary
problem.
Two observers are  shown falling through $r=r_+$ into the interior region and
then through the Cauchy horizon
at $r=r_-$. FFO1 (FFO2) crosses  the left (right) branch of $r=r_-$. An
incident wave is scattered
from the potential ($V^{OUT}$) outside the outer horizon. The scattered waves
by $V^{OUT}$ will be rescattered
into the hole, to give a tail with a power-law (in time) decay. Explicitly the
power-law tails come
from rescattering of the scattered waves off the weak potential far from the
black hole.
Since this decay rate  is sufficiently slow, it
 plays a  role to develop the infinite energy densities on the right branch
[11].
On the other hand, the transmitted wave proceeds into the interior region
where further scattering by  $V^{IN}$ occurs.  This scattered (right-moving)
wave is then
rescattered at $r^* \simeq 0$ to give a left-moving wave traveling near the $v
= \infty$ horizon.
Here  we investigate the way how the transmitted waves develope the infinite
energy density near
the $u=\infty$ branch.
In the non-extremal case ($Q^2<Q^2_e$), the tortoise coordinate is differently
given by
\begin{equation}
r^*= \int^r { dr \over f(r,\epsilon,Q)} = -{1 \over 2 \sqrt 2}\int^y { dy \over
yf(y,\epsilon,Q)},
\end{equation}
where  $f(y,\epsilon,Q)= 1-y + \tilde A y^{1+B}$ with $\tilde A
 = { Q^2 \over 4(2-\epsilon)}$.  Near  both horizons, $f(y,\epsilon,Q) \simeq
 f'(y_{\pm}) (y - y_{\pm}) $. Substituting this into (48), one finds
\begin{equation}
y- y_{\pm} \simeq  \pm \exp( \pm \kappa_{\pm} r^*),
\end{equation}
where $\kappa_{\pm}(r_\pm,\epsilon,Q) \equiv \pm f'(r_{\pm})(= \mp 2 \sqrt 2
f'(y_{\pm}) y_{\pm})$
are the surface gravitys at $r=r_{\pm}$. It is obvious from the graphical
analysis in
Fig.1 and Fig. 9 that $\kappa_{\pm}>0$.
 Near the outer horizon the potential decreases exponentially as
\begin{equation}
 V^{IN}_{r \to r_+}(r^*,\epsilon,Q) \simeq - 2 (2 f'(y_+) y_+ +1)f'(y_+) \exp
(\kappa_+ r^*),~~~
r^* \to -\infty  (r \to r_+)
\end{equation}
and near the inner horizon it takes the form
\begin{equation}
 V^{IN}_{r \to r_-}(r^*,\epsilon,Q) \simeq  2 (2 f'(y_-) y_- +1)f'(y_-) \exp
(-\kappa_- r^*),~~~
r^* \to \infty  (r \to r_-).
\end{equation}
We note that for the inner scattering analysis, the inner region ($r_-<r<r_+$)
should be changed
into $\infty>r^*>-\infty$.
It is useful to introduce the null coordinates ($ v = r^* + t, u = r^* - t$) to
describe the
inner structure of the charged black hole.  The metric is then given by $ds^2 =
fdvdu$.
As is shown in Fig.8, the Cauchy horizon $r=r_-$ consists of two branches
(the right with $v=\infty$ and the left with $u=\infty$).
In order to find the energy density of the tachyon measured by a freely falling
observer (FFO)
with two-velocity $U^{\mu} (U^{\mu}U_{\mu}=-1)$,
we have to consider the boundary conditions. Initially the tachyonic mode with
$\omega$ falls into the hole from
the exterior region. A general perturbation is a superposition of these
$\omega$ modes.
 Considering $t_\omega^{IN}(r^*,t) \sim t_\omega^{IN}(r^*) e^{-i\omega t}$,
one finds the equation for a particular frequency ($\omega$) near the horizons
\begin{equation}
\{ {\partial^2 \over \partial r^{*2}} + \omega^2 \} t^{IN}_{\omega}(r^*) = 0.
\end{equation}
Here we obtain  purely the ingoing wave near the event horizon $(r_{+})$
\begin{equation}
t^{IN}_{\omega} (r^*,t) \mid_{r_+} \equiv t^{IN}_{\omega}(u,v) \mid_{r_+}
 = T^{IN}(\omega) e^{-i\omega v}.
\end{equation}
On the other hand, the boundary condition near the Cauchy horizon is
\begin{equation}
t^{IN}_{\omega} (r^*,t) \mid_{r_-} \equiv t^{IN}_{\omega}(u,v) \mid_{r_-} =
e^{-i\omega v} +
 R^{IN}(\omega)e^{i\omega u},
\end{equation}
where the first term refers the ingoing mode into the left branch with
$u=\infty$, while the second
denotes the backscattered mode into the right branch with $v=\infty$.
Here $T^{IN}(\omega)$ and $R^{IN}(\omega)$ are the transmission and reflection
amplitudes for
the inner scatterings . These altered boundary conditions arise from the fact
that by virtue of
the light-cone structure of the inner region, one can have only ingoing modes
(by crossing the outer
horizon) and none leaving it (by crossing  the outer one  in the reverse
direction).
We take into account the general perturbation $(\widehat t^{IN}(r^*,t) = e^{-
\sqrt 2 r} t^{IN}(r^*,t) $
in (14)) to  obtain the  energy density. However, one has
$e^{- \sqrt 2 r} \sim e^{- \sqrt 2 r_-}$ near the Cauchy horizon
and thus can neglect it.
This is given by  the Fourier integral transfrom over the frequency $\omega$
\begin{equation}
\widehat t^{IN}(r^*,t) \sim  t^{IN}(r^*,t) = \int t^{IN}_{\omega}(r^*) e^{- i
\omega t} a(\omega) d\omega
\end{equation}
with the mode amplitude  $a(\omega)$.
Considering the boundary condition (54) near the Cauchy horizon, this takes the
form
\begin{equation}
t^{IN}(r^*,t)\mid_{r_-} = [t^{r}(v) + t^{l}(u)],
\end{equation}
where $t^r$ is a function of $v (= r^* + t)$ and $t^l$ is a function of $u (=
r^* - t)$.
The energy density measured by a FFO is dominated by [9]
\begin{equation}
\rho = T^t_{\alpha\beta}U^\alpha U^\beta \sim |U^{\alpha} t^{IN},_{\alpha}|^2
\sim |U^{\alpha}t^{l,r},_{\alpha}|^2.
\end{equation}
When  a FFO1 crosses the left branch ($u \to \infty$) of the Cauchy horizon,
one has
\begin{equation}
   U^{\alpha} t^{l},_{\alpha} \propto t^{l \prime}(u) \exp( {\kappa_- u \over
2}),
\end{equation}
where the prime means the differentiation with respect to the given argument.
In order to calculate $t^{l \prime}(u)$, we consider the deviation from the
wave
($t^{IN}_{\omega}(r^*)= \exp(-i\omega r^*)$)
treating $V^{IN}_{r \to r_-} \sim \exp (-\kappa_- r^*)$ in (51) as the
infinitesimal perturbation.
In this case one has a first-order scattering equation between two horizons :
$ (d^2/dr^{*2} +\omega^2)t^{IN}_\omega= V^{IN}_{r \to r_-} t^{IN}_\omega$.
It is a standard method to use Green's function in solving the above  equation.
Following Ref.[12], we find $t^{l \prime}(u) \propto \exp (-{\kappa_- u \over
2})$ as $u \to \infty$.
Therefore this wave gives a finite energy density at the left Cauchy horizon.
On the other hand, the energy density measured by a FFO2 who crosses the right
($v \to \infty$)
horizon is proportional to the square of
\begin{equation}
U^{\alpha} t^{r},_{\alpha} \propto t^{r \prime}(v) \exp( {\kappa_- v \over 2}).
\end{equation}
In order to calculate $t^{r \prime}(v)$, one  also consider the deviation from
the wave
($t^{IN}_{\omega}(r^*)= \exp(i\omega r^*)$)
treating $V^{IN}_{r \to r_+} \sim \exp (\kappa_+ r^*)$ in (50) as the
infinitesimal perturbation.
It is calculated  as $t^{r \prime}(v) \propto \exp (-{\kappa_+ v \over 2})$.
Substituting this  into (59)
leads to
\begin{equation}
U^{\alpha} t^{r},_{\alpha} \propto \exp \{ {(\kappa_- - \kappa_+) v \over 2}\}.
\end{equation}
{}From the graphical analysis in Fig.1 and Fig.9, it is easily shown that
$-f'(r_-)>f'(r_+)(
\kappa_->\kappa_+)$ for all $\epsilon<2, Q^2<Q_e^2$. This means that the slope
of the metric function $f$
at $r=r_-$ is always greater than the slope of $f$ at  $r=r_+$.
Thus (60) leads to a divergent
energy density on the right Cauchy horizon. This shows that the monochromatic
tachyon waves  with small amplitude
and purely ingoing near the event horizon in (53) develop the infinite energy
density near Cauchy horizon.
The FFO2 meets a divergent energy density when he (or she) crosses the right
branch of inner horizon.
This corresponds to the blueshift of tachyon. Further
this means that the Cauchy horizon of the 2D charged black hole is unstable to
the physical perturbations.

For the stability of outer horizon, one notes  the shape of the potentials
outside the
the outer horizon ($V^{OUT}$) in Fig. 5 to Fig. 7.  Outside the horizon
($r_+$),
all of the potentials take the positive barriers. This means that
the corresponding Schr\"odinger-type equations  allow only the scattering
states.
Since one cannot find any exponentially growing mode, the outer horizon of
non-extremal black hole
is stable.

\section{Hawking temperature}

The Hawking temperature of a static black hole can be calculated in several
ways [13].
Suppose that the metric takes the form
\begin{equation}
ds^2= -\lambda dt^2 + { dr^2 \over f}.
\end{equation}
Near the outer horizon $r=r_+$, one has $\lambda \simeq \lambda'(r_+)(r-r_+)$
and
$f \simeq f'(r_+)(r-r_+)$. We now set $\tau =it$ and $\rho = 2 \sqrt
{(r-r_+)/f'(r_+)}$.
The resulting metric is
\begin{equation}
ds^2=  d \rho^2 + { \lambda'(r_+) f'(r_+) \over 4}\rho^2 d \tau^2.
\end{equation}
In order to avoid a conical singularity at $\rho=0$ we must identify $\tau$
with period
$4\pi/ \sqrt {\lambda'(r_+) f'(r_+)}$. Thus the Hawking temperature is given by
\begin{equation}
T_H = {\sqrt {\lambda'(r_+) f'(r_+)} \over 4 \pi}.
\end{equation}
In our case, $\lambda =f$. Therefore the Hawking temperature takes the form
\begin{equation}
T_H = { |f'(r_+)| \over 4 \pi},
\end{equation}
which leads to
\begin{equation}
T_H = { e^{-2 \sqrt 2 r_+}  \over \sqrt 2 \pi} | {M \over \sqrt 2} - { Q^2
\over 8}({4-\epsilon \over 2-\epsilon})
e^{-(2-\epsilon) \sqrt 2 r_+}|.
\end{equation}
As is expexted, one finds  that  $T_H^e \to 0$ in the extremal limit of $Q^2
\to Q^2_e$ and
$r_+\to r_o$.
In the case of $\epsilon =0$, one finds the explicit form
\begin{equation}
T^Q_H = { \sqrt 2 \over \pi} { \sqrt{ M^2 -Q^2} \over (M + \sqrt{ M^2- Q^2})}.
\end{equation}
Further for the dilaton black hole ($\epsilon=Q=0$), the Hawking temperature
corresponds to the
statistical temperature
\begin{equation}
T^{Q \to 0,\epsilon \to 0}_H = { 1 \over \sqrt 2 \pi}.
\end{equation}

\section{discussions}
The 2D, $\epsilon<2$
extremal black holes have all zero Hawking temperature ($T^e_H=0$).
However one finds  the instability for 2D, $\epsilon<2$ extremal black hole,
which originates from
the  barrier-well potentials.
These potentials persist when the nonextremal black hole approaches the
extremal limit.
As is discussed in Ref.[14], the quantum stress tensor of a scalar field
(instead of the tachyon)
in the $\epsilon=0,$ 2D extremal black hole
diverges at the horizon. This means that the $\epsilon=0,$ 2D extremal black
hole is also
quantum-mechanically unstable.
This divergence can be better understood by the regarding an extremal black
hole as the limit of a
nonextremal one. A non-extremal black hole has an outer (event) and an inner
(Cauchy) horizon,
 and these come together
in the extremal limit.  In this case, it finds that if one adjusts  the quantum
state of the scalar
 field so that  the stress tensor is finite at the outer horizon, it always
diverges at
the inner horizon.
Thus it is not so surprising that in the extremal limit (when the two horizons
come together)
the divergence persists, although it has a softened form.
By the similar way, it would be conjectured that the classical instability of
$\epsilon <2$ extremal
black holes originates from the instability
(blueshift) of the inner horizon in the $\epsilon <2$ non-extremal black holes.
The potential of the non-extremal  black hole takes a barrier-well between the
inner and
 outer horizons,  while it takes a simple barrier outside the outer horizon. It
is confirmed that
the inner horizon is unstable, whereas the outer one is stable.
When these coalesce, a barrier-well type potential appears outside the event
horizon ($r>r_o$).
This induces the instability of the extremal black holes.
Contrary to the 2D calculations, the 4D extremal balck holes including
Reissner-Nordstr\"om one
 are all classically stable [3]. This is so because all potentials are positive
definite
outside the horizons. Further,  the quantum stress-energy tensor of a scalar
field
is recently calculated in the $a=0$  extremal Reissner-Nordstr\"om black hole
[15].
The stress-energy appears to be regular on the horizon. This establishes that
the $a=0,$ 4D extremal black hole is classically and quantum-mechanically
stable.

In conclusion, although the 2D  extremal charged black holes all have zero
Hawking
temperature, they  cannot be the toy models for the stable endpoint of the
Hawking evaporation.

\acknowledgments

This work was supported in part by Nondirected Research Fund, Korea Research
Foundation,1994
and by Korea Science and Enginnering Foundation, 94-1400-04-01-3

\figure{Fig. 1: Three graphs of the metric function $f(r,\epsilon,Q)$ for
$\epsilon =0.5$:
$Q^2>Q^2_e$ (dashed line : $-$-$-$-, no root);
$Q^2=Q^2_e$ (dotted line  :- - - -, extremal case with a multiple root );
 $Q^2<Q^2_e$ (solid line : ---, non-extremal case with two roots).

Fig. 2: Three graphs of extremal potentials ($V_e(r,\epsilon)$) for
$\epsilon = 1.9$ (dashed line : $-$-$-$-),
0.5 (dotted line  :- - - -), and $- 3$ (solid line : ---).
 The potentials are zero at $r_o(\epsilon) =-1.076, -0.299,$ and $-0.119$
and they are all barrier-well types outside $r_o(\epsilon)$.

 Fig. 3: The approximate  potential ($V^e_{app}(r^*,\epsilon)$).
The aymptotically flat region is at $r^* = \infty$.
This also takes a  barrier-well type. This is localized at $r^*=0$, falls to
zero
exponentially as $r^* \to \infty$ and inverse-squarely as $r^* \to -\infty$
(solid lines).
 The dotted line is used to connect two boundaries.

 Fig. 4: Three $\epsilon=0$ graphs of  potential for $Q=0.1$ (dashed line :
$-$-$-$-),
1 (dotted line  :- - - -), and $\sqrt 2$ (solid line : ---).
The corresponding event horizons are located at $r_+ = -0.004, -0.056$, and
$-0.245$ respectively.
When $M (= \sqrt 2) >Q (Q=0.1,1)$, the potentials outside the
event horizon are simple barriers. However  a barrier-well ($ V_e(r,\epsilon)$)
appears
as the nonextremal
black hole (a simple barrier) approaches the extremal one ($M=Q$).

 Fig. 5: $1\leq \epsilon<2$ graphs of the effective potential ($
V(r,\epsilon,Q)$) for $Q^2=Q^2_e/2$:
$\epsilon=1.9$ (dashed line : $-$-$-$-),
$\epsilon=1.6$ (dotted line  :- - - -), and $\epsilon=1.0$ (solid line : ---).
The potentials increase without bound in height as $\epsilon$ approaches its
upper limit (=2).

 Fig. 6: $ 0.1 \leq \epsilon<1$ graphs of the effective potential for
$Q^2=Q^2_e/2$ :
$\epsilon=0.8$ (dashed line : $-$-$-$-),
$\epsilon=0.5$ (dotted line  :- - - -), and $\epsilon=0.1$ (solid line : ---).
These take  the
   barrier-well type ($V^{IN}$) between two horizons, while it takes a simple
potential
 barrier ($V^{OUT}$) outside the outer horizon.

 Fig. 7: $ \epsilon<0$  graphs of the effective potential for $Q^2=Q^2_e/2$ :
$\epsilon=-0.5$ (dashed line : $-$-$-$-),
$\epsilon=-3$ (dotted line  :- - - -), and $\epsilon=-10$ (solid line : ---).
These take also the
   barrier-well type ($V^{IN}$) between two horizons, while it takes a simple
potential
 barrier ($V^{OUT}$) outside the outer horizon. As $\epsilon$ decreases,
 the inner potentials are shifted to $V$-axis. And the shapes of outer
potentials are not significantly changed. .

 Fig. 8: Conformal diagram of a portion of the 2D non-extremal  charged black
hole space-time.
Two observers are  shown falling through $r=r_+$ into the interior region and
then through the Cauchy horizon
at $r=r_-$. FFO1 (FFO2) crosses  the left (right) branch of $r=r_-$. An
incident wave is scattered
from the potential ($V^{OUT}$).  The scattered wave by $V^{OUT}$ will be
rescattered into hole, to give a tail with a
power-law (in time) decay. This decay in time  is
sufficiently slow so that infinite energy densities are developed on the right
branch [11].
On the other hand, the transmitted wave proceeds into the interior region where
further scattering
by  $V^{IN}$ occurs. This right-moving wave is
rescattered at $r^* \simeq 0$ to give a left-moving wave traveling near the $v
= \infty$ horizon.
Here  we show the way that these waves develope the infinite energy density
(measured by FFO2)
near the $u=\infty$ branch.

 Fig. 9: Three graphs of non-extremal $f(r,\epsilon,Q)$ for $Q^2=Q^2_e/2$ :
$\epsilon =1.5$ (dashed line : $-$-$-$-),
0.1 (dotted line  :- - - -), and $-3$ (solid line : ---). It is obvious that
$-f'(r_-)>f'(r_+)$
for all $\epsilon<2$. As $\epsilon$ approaches 2, the decreasing rate of $f$ at
the inner horizon(
 $r=r_-$) is always greater than the increasing rate of $f$ at the outer
horizon ($r=r_+$).
}

\end{document}